\magnification1200

\rightline{KCL-MTH-02-31}
\rightline{hep-th//0212291}

\vskip .5cm
\centerline
{\bf Very Extended  $E_8$ and $A_8$ at low levels,    
Gravity and Supergravity  }
\vskip 1cm
\centerline{  Peter West }
\vskip .5cm
\centerline{Department of Mathematics}
\centerline{King's College, London, UK}

\leftline{\sl Abstract}
\noindent 
We define a  level for a large class of Lorentzian
Kac-Moody algebras. Using this we find the representation content of 
very extended $A_{D-3}$ and $E_8$ (i.e. $E_{11}$) at low levels in terms
of $A_{D-1}$ and 
$A_{10}$ representations respectively.  The results are consistent  with
the conjectured  very extended $A_8$ and $E_{11}$ symmetries of gravity
and maximal supergravity theories given respectively in hep-th/0104081 and
hep-th/0107209. We explain how these results   provided further
evidence  for these conjectures. 
 
\vskip .5cm

\vfill

\vskip 1cm
email:  pwest@mth.kcl.ac.uk

\eject


\medskip
{\bf {1. Introduction }}
\medskip
One of the most surprising discoveries 
in supergravity theories was the realisation that the maximal
supergravity theory in four dimensions possessed a hidden $E_7$ symmetry
[1]. Hidden symmetries were found in the other maximal supergravity
theories, and the lower the dimension of the maximal supergravity theory
the higher the rank of the hidden symmetry and the more interesting it
was.  An account of these, with their original references, can be found in
[2].  Over the years there have also been some other realisations of
possible symmetry algebras in supergravity and string theories. 
 It has been shown [3,4] that eleven dimensional
supergravity  does possess an SO(1,2)$\times$SO(16) symmetry,
although the SO(1,10) tangent space symmetry is
no longer apparent in this formulation.  It has also been noticed 
that some of the
objects associated with the exceptional groups that  appear in the
reductions  appear naturally in the unreduced theory [5].  In string
theory it was found that  the closed bosonic
string reduced on the torus associated with  the unique self-dual twenty
six dimensional Lorentzian lattice is invariant under
the fake monster algebra [6]. It has also been suggested [7] that there is
some evidence of
Kac-Moody structures in threshold correction of the heterotic string
reduced on a six dimensional torus.  Finally,  it has been
found that the maximal supergravity theory in eleven dimensions, in a
special limit, possess one dimensional solutions  which
correspond to  a particle mechanics in which the  motion is restricted to
the Weyl chambers of
$E_{10}$ [8]. 
\par
Despite these observations it  has been widely 
thought that the exceptional groups found in the dimensional
reductions of the maximal supergravity theories can not be  symmetries
of eleven dimensional supergravity and must arise as a consequence of the
dimensional reduction procedure.  This is  perhaps understandable given
that the hidden symmetries were associated with scalar fields and there
are no scalar fields in eleven dimensional supergravity. However, 
 it has been conjectured
[2] that  the eleven dimensional  supergravity theory possess a hidden
$E_{11}$ symmetry.  It had previously been found [9]
that the whole of the bosonic sector of eleven dimensional supergravity
including its gravitational sector was a non-linear realisation. This
placed the fields of the theory on a equal footing and naturally
incorporated symmetries even though there were no scalars. This
construction contained substantial fragments of larger symmetries and
suggested that these should be incorporated into a Kac-Moody, or more
general  symmetry group. Even though it was not shown that such a
symmetry was realised  one could determine that this symmetry should
contain  very extended $E_8$, i.e. 
$E_{11}$ [2].  It was also  shown that this construction could be
generalised to the ten dimensional IIA [2] and IIB [10] supergravity
theories and the corresponding Kac-Moody algebra was also $E_{11}$ in each
case. It was also proposed that gravity in D dimensions should also have 
a hidden Kac-Moody symmetry which was very extended $A_{D-3}$ [11]. 
\par
These ideas were taken up  by  authors of reference [12] who 
considered  eleven dimensional supergravity as a
non-linear realisation of  the
$E_{10}$  subalgebra of
$E_{11}$, but in the small tension limit
which played a crucial role in the work of reference [8]. These authors
also introduced   a new concept of level in the context of
$E_{10}$ which allowed them to deduce the representation content of
$E_{10}$ in terms of representations of
$A_9$ at low levels. They showed that the eleven
dimensional supergravity equations was  $E_{10}$ invariant, 
up to level three,  
 in the small tension limit and provided one adopted a particular map
relating the fields at a given spatial point to quantities dependent
only on time.  In this limit the spatial dependence of the fields was
very restricted. 
\par
In this paper we generalise the notion of level given in reference [12]
for a large class of Lorentzian Kac-Moody algebras. In section three we 
calculate the 
 representation content of $E_{11}$ and very extended $A_{D-3}$ at low
levels in terms of $A_{10}$ and $A_{D-1}$ representations respectively. 
We then apply these results in sections four and five to the conjectures
that gravity in D dimensions in invariant under very extended $A_{D-3}$
and eleven dimensional supergravity is invariant under $E_{11}$
respectively.   
\medskip
{\bf {2. Levels in Lorentzian Kac-Moody Algebras}}
\medskip
We consider the particular class of Kac-Moody algebras discussed in
section three of reference [13], namely an algebra whose Dynkin
diagram $C$ possess at least one   node such that deleting it leads to a
finite dimensional semi-simple Lie algebra. We denote the Dynkin diagram
of this remaining Lie algebra by  
$C_R$. For simplicity, we will consider the case when this remaining
algebra is irreducible, but the generalisation is
obvious. Adopting the notation of reference [13]; we denote the
preferred simple root by
$\alpha_c$ and the simple roots of $C_R$ by $\alpha_i, i=1,\ldots , r-1$.
these may be written as [13]   
$$\alpha_c=-\nu +x    
\eqno(2.1)$$
where $\nu=-\sum_i A_{ci}\lambda_i$, $x$ is a vector in a space
orthogonal to the roots of $C_R$ and $\lambda_i$ are the fundamental weight
vectors  of
$C_R$. We will assume that the simple roots have length squared two and
then $2-\nu^2=x^2$. 
\par
A positive root $\alpha$ of the Kac-Moody algebra $C$ can be written as
$$\alpha= l\alpha_c +\sum_im_i\alpha_i
=lx+l\sum_i A_{ci}\lambda_i+\sum_{jk} A^f_{jk}\lambda_k
\eqno(2.2)$$
where $A^f_{jk}$ is the Cartan matrix of $C_R$.
We define the level, denoted $l$, of the roots of $C$ to be the number
of times the root $\alpha_c$ occurs. This agrees with the notion
of level introduced in reference [12]  for the case of $E_{10}$. 
For a fixed
$l$, the root space of
$C$ can be described by its representation content with respect to $C_R$.
We now 
discuss restrictions that one can place on the possible
representations that can occur at a given level generalising the
considerations of reference [12] for $E_{10}$. 
Given  an irreducible  representation with highest
weight
$\Lambda$ it is completely specified by its Dynkin indices which
 are defined by 
$p_j=(\Lambda,\alpha_j)\ge0$. It follows from equation (2.2) that a 
representation of $C_R$, with Dynkin indices $p_j$, is contained in $C$ at
level
$l$ if  there exist positive integers $m_k$ such that 
$$ p_j=lA_{cj} +\sum_k m_k A^f_{kj}
\eqno(2.3)$$
\par
Any Kac-Moody algebra with symmetric Cartan matrix has its roots bounded
by $\alpha^2 \le 2,1,0\ldots $ [14]. Assuming we are dealing with such a
Kac-Moody algebra,   using
equations (2.1) and (2.2), and the observation that
$(2-\nu^2)
\det A_{C_R}=A_{C}$ we find that 
$$\alpha^2=l^2 {\det A_C\over \det A_{ C_R}}+\sum_{ij}p_i (A^{f
}_{ij})^{-1} p_j\le 2,1,0,-1,\ldots
\eqno(2.4)$$
Since the first term is fixed for a given $l$ and the second term is
positive definite this places constraints on  the possible values of
$p_i$. 
\par
Acting  with   the raising or lowering operators of 
$C_R$ on the adjoint representation of $C$ takes one from one  root of
$C$  to a new root of
$C$. The original and the new roots of $C$ are  of the form of equation
(2.2) with
$l$ and
$m$ positive or negative integers depending if the root is in the
positive of negative root space,  However, under the action of such
operators   the  first term of this equation, i.e.
$lx$, is  unchanged, and so the level of the new root is the same. As the
positive or negative roots have all their coefficients of a given sign we
 find that   
 the action of these raising or lowering operators does not take one out
of the positive or negative  root space of $C$. As a result, we may
conclude that  if a given
representation of $C_R$ occurs in $C$ then it must be contained entirely
in the positive roots with a copy in the  negative root space of $C$. 
By
considering a negative root, which has the form of equation (2.2), but
with a negative sign on all terms on the right hand-side, we conclude
that $C$ may contain the representation of $C_R$ with Dynkin indices
$p_j$ if 
$$\sum_j (A^{f }_{kj})^{-1} p_j=-l\sum _j A_{cj} (A^{f }_{jk})^{-1}-m_k
\eqno(2.5)$$ 
where $m_k=0,1,2,\ldots $.  Since the left-hand side  and the first term
on the right-hand side are both positive, one finds, for fixed $l$, 
constraints on the allowed representations of
$C_R$ that can occur.
\par 
A solution to the constraints of equations
(2.4) and (2.5) implies that the corresponding  representation of $C_R$
can occur in the root space of 
$C$, but it does not imply  that it actually does occur. The above the
constraints find the vectors   in the  lattice spanned by the
positive or negative roots of $C$ that have length squared
$2,1,0,-1,\ldots$ and contain highest weight vectors of
representations of the reduced subalgebra $C_R$. This is not the same  as
starting from the simple roots, taking their multiple commutators and
imposing the Serre relations. By definition the latter calculation leads
to all the roots of the Kac-Moody algebra and no more . However, given the
list of possible representation that satisfy the constraints of equations
(2.4) and (2.5) we can consider the set of associated generators  and 
 construct the Kac-Moody algebra $C$ up to the level
investigated by insisting that it satisfies all the consequences of the
Serre relations. One simple requirement is that they satisfy the Jacobi
identities. In some of the examples considered below one finds that  this
step implies that some of the potential representations of the reduced
algebra do not actually  belonging to the Kac-Moody algebra. 
In particular, we will see that  some  vectors contained in the 
root lattice which do satisfy the constraints  are actually highest
weight vectors of the Kac-Moody algebra $C$ and do not actually belong to
the  Kac-Moody algebra. 
Indeed, the criterion given above does not distinguish between the
Kac-Moody algebra $C$ and the algebra derived from the physical
state vertex operators associated with the roots lattice. As the latter
contains many more states than the Kac-Moody algebra one expects to find 
these additional states by using the criterion above. 
\par
Finding the root multiplicities of Kac-Moody algebras is an unsolved
problem except for a few exceptional cases. One may hope that the concept
of level given above may provide a new avenue of attack on this problem. 
\medskip 
{\bf {3. Representations of very extended $E_8$ and $A_{D-3}$ at low
levels}}
\medskip
In this section we find the representation content of  very extended
$E_8$ and $A_{D-3}$ at low levels in terms of representations of
$A_{10}$ and $A_{D-1}$ respectively. Let us begin with  very extended
$E_8$, i.e. 
$E_{11}$. The Dynkin diagram for this algebra is found by
connecting   ten dots in  a horizontal line 
 by a single line and then placing another dot above
the third node from the right and connecting it with a single line to
that node. The  preferred node, c is the  node above the horizontal line
and deleting it gives
$A_{10}$. Clearly,  at level zero  we have the adjoint representation of
$A_{10}$.  
The inverse Cartan matrix of $A_{D-1}$ is
given by 
$$(A^{f }_{jk})^{-1}=\cases{{j(D-k)\over 11}, \ \ j\le k\cr
{k(D-j)\over D}, \ \ j\ge k\cr}
\eqno(3.1)$$
Using this fact for $D=11$,  the constraint of equation (2.5) becomes 
$$ \sum_{j\le k}j(11-k)p_j+\sum_{j> k} k(11-j)p_j=-11n_k+l
\cases{3(11-k),\ k\ge 3\cr 8k,    \ \ \ \ \ \ \ k=1,2} ,
\eqno(3.2)$$ 
where  $n_k=0,1,2\ldots$. Analysing this equation and equation
(2.4) using the relations  $\det A_{E_n}=9-n$ and  $\det A_{A_n}=n+1$,
it is straightforward, if tedious, to verify that they 
 allow  the following representations of $A_{10}$ 
$$ l=1,\ p_3=1\ \ ;
l=2,\ p_6=1\ \ ;
l=3,\ p_1=1,p_8=1\ {\rm and}\ p_9=1\ \ ;
$$
$$l=4,\ p_{10}=1,p_1=2\ {\rm and}\  p_{10}=1,p_2=1\ {\rm and}\ 
p_9=1,p_3=1 \ {\rm and}\ p_1=1
\eqno(3.3)$$
All other $p_j$'s being zero [20]. 
\par
The corresponding   generators  are   given by 
$$K^a{}_b, R^{a_1a_2a_3}, R^{a_1a_2\dots a_6},R^{a_1a_2\ldots a_8,b}
\ {\rm and }\ R^{a_1a_2\ldots a_9},
\eqno(3.4)$$
where 
$R^{[a_1a_2\ldots a_8,b]}=0$, at levels zero, one, two and three
repectively.   
These generators are precisely those introduced in the
non-linear realisation of eleven dimensional supergravity based on
$E_{11}$ [2] with the exception of the last generator,  
the $p_9=1$ representation at level three, which is absent from
the $E_{11}$ algebra as a result of the Jacobi identities[2]. 
They obey the commutation relations [2]
$$
[K^a{}_b,K^c{}_d]=\delta _b^c K^a{}_d - \delta _d^a K^c{}_b,  
\eqno(3.5)$$
$$  [K^a{}_b, R^{c_1\ldots c_6}]= 
\delta _b^{c_1}R^{ac_2\ldots c_6}+\dots, \  
 [K^a{}_b, R^{c_1\ldots c_3}]= \delta _b^{c_1}R^{a c_2 c_3}+\dots,
\eqno(3.6)$$
$$[ R^{c_1\ldots c_3}, R^{c_4\ldots c_6}]= 2 R^{c_1\ldots c_6},\ 
\ \ 
[R^{a_1\ldots a_6}, R^{b_1\ldots b_3}]
= 3  R^{a_1\ldots a_6 [b_1 b_2,b_3]}, 
\eqno(3.7)$$
$$ [ R^{a_1\ldots a_8, b} ,R^{b_1\ldots b_3}]=0, \ 
[ R^{a_1\ldots a_8, b} ,R^{b_1\ldots b_6}]=0, \ 
 [ R^{a_1\ldots a_8, b} ,R^{c_1\ldots c_8,d}]=0
\eqno(3.8)$$ 
$$ [ K^a{}_b,  R^{c_1\ldots c_8, d} ]= 
(\delta ^{c_1}_b R^{a c_2\ldots c_8, d} +\cdots) + \delta _b^d
R^{c_1\ldots c_8, a} .
\eqno(3.9)$$
The full $E_{11}$ algebra up to level three is then given by these
relations plus those involving the negative roots 
$$ R_{a_1a_2a_3}, R_{a_1a_2\dots a_6},R_{a_1a_2\ldots a_8,b}
\eqno(3.10)$$
The existence of the generators up to level three can also  be inferred by
the considerations of U-duality groups in the toroidal reduction of eleven
dimensional supergravity [19]. 
\par
As explained in the previous section, it can happen that there
exist vectors in the root lattice that are solutions of the constraints,
but are  actually  highest weight
vectors of the Kac-Moody algebra. The fundamental weights  of the
Kac-Moody algebra are given by [13]
$$ l_i=\lambda_i+\nu.\lambda_i{x\over x^2},\  l_c={x\over x^2}
\eqno(3.11)$$
The $p_9=1$ representation at level three of $E_{11}$ corresponds to the 
vector $-3x+\lambda_9$. Using equation (3.11) we recognise this vector as
just the fundamental  weight $l_9=-3x+\lambda_9$ of $E_{11}$. Constructing
the root string of $l_9$ we find that it contains the vector 
$$l_9-\alpha_9-\ldots -\alpha_3-\alpha_c=-4x+\lambda_2+\lambda_{10}, 
\eqno(3.12)$$
where in the last equation we have re-expressed the vector in the root
string in terms of
$x$ and fundamental weights of $A_{10}$. Clearly, the  highest weight
representation of $E_{11}$ contains at level four the $p_2=1,p_{10}=1$,
all other $p_j$'s zero, representation of $A_{10}$. Examining the
solutions of the constraints at level four of equation (3.3) we recognise
that this is one of the representations that occurs and so it must also
be eliminated from the $E_{11}$ algebra.  
\par
We now consider very extended
$A_8$. Its Dynkin diagram is given by drawing 10 dots on a horizontal line
and  joining them together by a single line, we then draw a node above the
horizontal line and  join it with a single line to the node on the far
right and the third node from the left.   The preferred node, labeled as
$c$, is the one above the horizontal. Deleting this node leaves the nodes
in the horizontal line which correspond to the algebra
$A_{10}$.  At level zero one has the
adjoint representation of $A_{10}$. For this case equation (2.5) becomes 
$$ \sum_{j\le k}j(11-k)p_j+\sum_{j> k} k(11-j)p_j=-11m_k+l
\cases{(11+2k),\ k\le 8\cr 9(11-k),    \ \ \ \ \ \ \ k=9,10} ,
\eqno(3.13)$$ 
where $m_k=0,1,2\ldots$. Analysing this equation and 
 equation (2.4)  implies that at the first two
levels one has the potential representations 
$$ l=1,\ p_1=1,p_8=1\ {\rm and}\  p_9=1\ \ ;
$$
$$l=2,\ p_{1}=1,p_6=1,\ {\rm and}\ 
p_{2}=1,p_5=1\ {\rm and}\  p_{8}=1,p_{10}=1,\ {\rm and}\ 
p_{1}=1,p_7=1,p_{10}=1
\eqno(3.14)$$
all other $p_j$'s zero. Hence we have the generators 
$$K^a{}_b, R^{a_1\ldots a_8,b}
\eqno(3.15)$$ 
where $R^{[a_1\ldots a_8,b]}=0$, at levels zero and one. We have not
included a generator $ R^{a_1\ldots a_9}$ since we will explain below
that this is not actually in the very extended $A_8$ algebra. 
\par
In fact,  very extended $A_8$ is a subalgebra of $E_{11}$ since 
$A_8$ is a subgroup of $E_8$. As a result,  all the $A_{10}$
representations that  occur as representations of very extended $A_8$ 
must also 
be representations of $E_{11}$. A detailed examination of  the
constraints of equations (2.4)  applied to $A_{10}$ and
$E_{11}$ reveals that a
$A_{10}$  representation
of very extended $A_8$ at level $l$ will be a representation of $E_{11}$ 
at level $3l$. Furthermore, equation  (2.5) becomes equations (3.2) 
and   (3.13) for the cases of $E_{11}$ and 
$A_{10}$ respectively and  we find that 
if we take the level in the former equation to be $3l$ where $l$ is the
level of very extended $A_8$ then equation (3.13) becomes equation (3.2) 
provided  we identify  
$$n_k=\cases{m_k +l(2k-1),\ k=1,2\cr 
m_k +l(8-k),\ k=3,\dots,8\cr
m_k,\ k=9,10}
\eqno(3.16)$$
Since if $m_k$ is a positive integer so is   $n_k$, we find that
solutions to equation (3.13) at level $l$ are solutions of equation
(3.2) at level $3l$ as required. Indeed, one sees that the 
representations of equation (3.14)  at level one occur in equation (3.3)
at level three and one can explicitly check that the representations at
level two of very extended $A_8$ occur as solutions of the
$E_{11}$ equations at level six.  
\par
Finally, we consider the case of very extended $A_{D-3}$ which
generalises the previous case to arbitrary rank. 
Its Dynkin diagram is given by drawing $D-1$ dots on a horizontal line
and  joining them together by a single line, we then draw a node above the
horizontal line and  join it with a single line to the node on the far
right and the third node from the left.   The preferred node, labeled as
$c$, is the one above the horizontal. Deleting this node  leaves the
nodes in the horizontal line which correspond to the algebra
$A_{D-1}$. 
At level zero one has the
adjoint representation of $A_{D-1}$ with generators $K^a{}_b$.  Analysing
equations (2.4) and 
 equation (2.5)  implies that at 
level  one has the potential representations 
$$ l=1,\ p_1=1,p_{D-3}=1\ {\rm and}\  p_{D-2}=1 ,
\eqno(3.17)$$
all other $p_j$'s zero. Hence, we have the generators 
$$K^a{}_b, R^{a_1\ldots a_{D-3},b}
\eqno(3.18)$$ 
where $R^{[a_1\ldots a_{D-3},b]}=0$,
at levels zero and one. We have not included a generator $R^{a_1\ldots
a_{D-2}}$ since, as we will explain below, this is not actually in the
very extended $A_{D_3}$ algebra. To derive  this result we first realise
that equation (2.5) implies results for
$k=1$ and
$k=D-1
$ which, when added together, imply that 
$$\sum_{j=1}^{D-1} p_j=2l, 2l-1,\ldots
\eqno(3.19)$$
Hence, at level one one finds that at most two $p_j$'s can be non-zero.
Equation (2.5) for $k=D-1$ implies that at level one $p_{D-1}=0$
and  the $k=1$ constraint implies that if $p_1=1$ then the only solution
 has $p_{D-3}=1$ at level one. To look for the other possible 
solutions we may take $p_1=p_{D-1}=0$,  then adding the $k=2$
constraint to the $k=D-2$ constraint, one finds that 
$$2\sum_{j=2}^{D-2}  p_j=3l, 3l-1,\ldots .
\eqno(3.20)$$
Hence, at level one at most one of the remaining $p_j$'s can be non-zero.
Examining the remaining constraints one finds that the only remaining
solution is that listed in equation (3.17). 
One can show that equations similar to (3.19) and (3.20) can also be
derived  for the case of 
$E_{11}$
\par
Just  like  the case of $E_{11}$, the solution $p_{D-2}=1$ of equation
(3.17) is in fact not a representation that occurs in the very extended  
$A_{D-3}$ Kac-Moody algebra. The corresponding vector in the root lattice
is
$-x+\lambda_{D-2}$ and, following the same arguments as above, we find
that  this vector is just the highest weight vector of the 
  representation   of very extended $A_{D-3}$ with
fundamental highest weight $l_{D-2} =-x+\lambda_{D-2}$. One finds that at
level two this representation contains the vector 
$-2x+\lambda_{D-4}+\lambda_1+\lambda_{D-1}$ which is the $
p_{D-4}=1,p_{D-1}=1,p_1=1$ representation. One can verify that at level
two this is a possible solution and so this solution   must also be
excluded from the very extended 
$A_{D-3}$ algebra. 
\par
The generators of very extended $A_{D-3}$ given in equation (3.18) obey
the commutation relations
$$ [K^a{}_b,K^c{}_d]=\delta _b^c K^a{}_d - \delta _d^a K^c{}_b,  
\eqno(3.21)$$
$$ [ K^a{}_b,  R^{c_1\ldots c_{D-3}, d} ]= 
(\delta ^{c_1}_b R^{a c_2\ldots c_{D-3}, d} +\cdots) + \delta _b^d 
R^{c_1\ldots c_{D-3}, a} .
\eqno(3.22)$$
The
complete  very extended
$A_{D-3}$ algebra up to level one is then given by the above relations
plus those involving  the negative root generators 
$$R_{c_1\ldots c_{D-3},b}.
\eqno(3.23)$$
\medskip 
{\bf {4. Dual Gravity and Very extended $A_{D-3}$}}
\medskip 
The bosonic sector of eleven dimensional supergravity, including its
gravitational sector,  was shown [9] to be a non-linear realisation. It
was subsequently, realised that this formulation possessed the Borel
subgroup of
$E_7$ as a symmetry and it was conjectured that this theory could be
formulated in such a way that it was  invariant under
$E_{11}$ [2]. 
Demanding that the $E_8$ borel subgroup and the
$A_{10}$ subgroup be symmetries implied  that one must
reformulate the   theory  so that gravity was described 
  by a dual formulation involving the fields
$h_a{}^b$ and  
$h_{a_1a_2\ldots a_{D-3},b}$ [2].  An action for a 
 dual theory of gravity in
$D$ dimensions was given [2] which was equivalent to Einstein's
theory, it contained the vierbein
$e_\mu ^a=(e^{h})_\mu{}^a$  and a field  $Y_{a_1\ldots
a_{D-2}, b}$.  
The corresponding   equations of motion were given by 
$$ (-1)^{D-2}{1\over (D-3)!}\epsilon^{\mu\nu\tau_1\ldots \tau_{D-2}} 
Y_{\tau_1\ldots \tau_{D-2},} {}_d
=2e(\omega_d, {}^{\mu\nu}+e_d{}^\nu \omega_c, {}^{c}{}^{\mu}-
e_d{}^\mu{} \omega_c, {}^{c}{}^{\nu})
\eqno(4.1)$$
where $e=\det e^\mu{}^a$, $ \omega_\rho,{}_{c}{}^d$ is the usual
expression for the  spin connection in terms of the vierbein, and 
$$ \epsilon^{\mu\tau_1\ldots \tau_{D-1}} \partial_{\tau_1}
Y_{\tau_2\ldots \tau_{D-1},} {}_d={\rm terms \ bilinear\  in}\  Y
\eqno(4.2)$$
The field $Y_{\tau_1\ldots \tau_{D-2},} {}_d$ has been scaled by a
numerical factor compared to that of reference [2]. 
\par  
At the
linearised level, equation (4.2) implies that $Y_{\tau_1\ldots
\tau_{D-2},} {}_d$ can be solved in terms of $ h_{\tau_2\ldots
\tau_{D-2}}, b$.  Substituting this
into   equation (4.1) and writing $\omega_{\mu,cd}$ in terms of
its standard linearised expression in terms of $e_\mu{}^b$ one
finds the   equation
$$
(-1)^{D-2}{(D-2)\over (D-3)!}\epsilon^{\mu\nu\tau_1\ldots
\tau_{D-2}}\partial_{[\tau_1} h_{\tau_2\ldots
\tau_{D-2}], d}= -\partial_\mu (e_{d\nu}+e_{\nu d})+\partial_d
(e_{\mu\nu}+e_{\nu \mu}) +\partial_\nu (e_{d\mu}+e_{\mu d})
$$
$$+2\eta_{\nu d}(\partial_\mu e_{c}{}^c
-\partial_c e_\mu{}^c)-2\eta_{\mu d}(\partial_\nu e_{c}{}^c
-\partial_c e_\nu{}^c)
\eqno(4.3)$$. 
Taking $\partial_\mu$ implies the 
linearised Einstein equation. Carrying out a local Lorentz transformation 
$\delta e_{\mu\nu}=\lambda _{\mu\nu}$ one finds it is a symmetry provided 
$\delta h_{\tau_2\ldots\tau_{D-2}, d}=-\epsilon_{\tau_2\ldots
\tau_{D-2} dm\rho\kappa}\lambda ^{\rho\kappa}$. 
\par
Based on these, and other 
considerations,  it was proposed [11] that even
pure gravity  in D dimensions could be described as  a
non-linear realisation based on very extended $A_{D-3}$.    
The results of section three provide encouraging signs for this
conjecture. Since eleven dimensional supergravity contains gravity, any
conjecture for the Kac-Moody symmetries of gravity and eleven dimensional
supergravity  theories would have to be consistent.  As explained in
section three, this is the case since 
$E_{11}$ 
 contains very extended
$A_{8}$ as a sub-algebra.
Furthermore, the very extended $A_{D-3}$ given  in
equation (3.17) contains the generators $K^a{}_b$ and $ R^{a_1\ldots
a_{D-3},b}$ at levels zero and  these imply that the non-linear
realisation is built from the fields $h_a{}^b,\hat  h_{a_1\ldots
a_{D-3},b}$ plus fields which appear at higher levels.  The level zero
and one fields  are almost exactly those used in the above dual
formulation of gravity. The difference is that the very extended $A_8$
algebra requires that 
$ R^{[a_1\ldots a_{D-3},b]}=0$ with a corresponding contraint for  the
field $\hat  h_{a_1\ldots a_{D-3},b}$
\par
It is instructive to derive Einstein's equation from the 
field equations rather than the action, as was the case in reference [2].
Since some of the manipulations are best carried out in form language and
some in components we given both versions. We first consider the tensor   
$$ Y_{\rho_1\ldots \rho_{D-2},}{}_a=\epsilon_{a b_1\ldots
b_{D-1}}e_{[\rho_1}{}^{b_1}\dots e_{\rho_{D-3}}{}^{b_{D-3}} 
\omega_{\rho_{D-2}]}{}^{b_{D-2}b_{D-1}} ,
\eqno(4.4)$$ whose corresponding  vector valued $D-2$
form is given by 
$$ Y_a=\epsilon_{a b_1\ldots b_{D-1}}e^{b_1}\wedge\dots \wedge e^{b_{D-3}}
\wedge w^{b_{D-2} b_{D-3}}
\eqno(4.5)$$
where $e^{a}=dx^\mu e_\mu {}^a$ and $w^{b c}=dx^\mu
w_\mu{}^{b c} $. 
The equation of motion is given by 
$$\epsilon ^{\mu\nu\rho_1\ldots \rho_{D-2}}(\partial_\nu Y_{\rho_1\ldots
\rho_{D-2},}{}_a-M_{\nu\rho_1\ldots \rho_{D-2},a})=0,\ \ {\rm or}\ \ 
dY_a=M_a.
\eqno(4.6)$$
where the vector valued $D-1$ form is defined by 
$$M_{\rho_1\ldots \rho_{D-1},a}=-\epsilon_{a b_1\ldots
b_{D-1}}\{ (D-3) w_{[\rho_1}{}^{b_1}{}_f e_{\rho_2}{}^f
e_{\rho_{2}}{}^{b_2}\dots
e_{\rho_{D-2}}{}^{b_{D-3}}w_{\rho_{D-1}]}{}^{b_{D-2}b_{D-1}}
$$
$$
+(-1)^{(D-3)} e_{[\rho_{1}}{}^{b_1}\dots e_{\rho_{D-3}}{}^{b_{D-3}}
w_{\rho_{D-2}}{}^{b_{D-2} f} w_{\rho_{D-1} ]f}{}^{  b_{D-1}}\}
\eqno(4.7)$$ 
or 
$$M_{a}=-\epsilon_{a b_1\ldots
b_{D-1}}\{ (D-3) w^{b_1}{}_f \wedge e^f\wedge 
e^{b_2}\wedge \dots\wedge 
e^{b_{D-3}}\wedge w^{b_{D-2}b_{D-1}}
$$
$$
+(-1)^{(D-3)} e^{b_1}\wedge \dots \wedge e^{b_{D-3}}\wedge 
w^{b_{D-2} f} \wedge w_{ f}{}^{  b_{D-1}}\}
\eqno(4.8)$$
Evaluating equation (4.6) we find that it 
becomes 
$$
\epsilon_{a b_1\ldots b_{D-1}}\epsilon ^{\mu\nu\rho_1\ldots \rho_{D-2}}
\{{(D-3)\over 2}T_{\nu\rho_1}{}^{b_1}e_{\rho_{2}}{}^{b_2}\dots
e_{\rho_{D-3}}{}^{b_{D-3}} w_{\rho_{D-2}}{}^{b_{D-2}b_{D-1}}\}
$$
$$-2\det e (D-3)!(-1)^{D-2}(R_a{}^\mu-{1\over 2}e_a{}^\mu R)=0
\eqno(4.9)$$
In these equations the torsion  and Riemann tensors 
are given by 
$$T^a=T_{\mu\nu}{}^a dx^\mu\wedge  dx^\nu=2(d e^a+w^a{}_b\wedge e^b),
$$
$$
 \ R^a{}_b=R_{\mu\nu}{}^a{}_b  dx^\mu \wedge dx^\nu= 2
(d w^a{}_b+w^a{}_c\wedge w^c{}_b ),\
R_{\mu\nu}{}^\mu{}_b=R_{\nu}{}_b
\eqno(4.10)$$
The first term in equation (4.9)  vanishes as 
the spin connection has the usual expression
in terms of the vierbein and so the torsion tensor
$T_{\mu\nu}{}^a$ vanishes. Hence, we are left with the familiar equation
for general relativity without matter. The above formulation of
general relativity agrees with that given in reference [18] for four
dimensional space-time where a connection with twistor theory was made
that may prove useful in future. 
\par
The equation of motion of equation (4.6) is second order in space-time
derivatives, but the above equations are written in a such way that they 
allows us to express them as a system of equations that is first order
in space-time derivatives by introducing the fields  
$$e_\mu ^a=(e^{h})_\mu{}^a,  h_{a_1\ldots a_{D-3},b}
\ {\rm and}\ k_{a_1\ldots a_{D-2},b}
\eqno(4.11)$$
The equations of motion are now given by  
$$\epsilon ^{\mu\nu\rho_1\ldots \rho_{D-2}}(
Y_{\rho_1\ldots \rho_{D-2}}{}_a -k_{\rho_1\ldots
\rho_{D-2}, a})
=\epsilon ^{\mu\nu\lambda\rho_1\ldots \rho_{D-2}}\hat D_\lambda
h_{\rho_1\ldots
\rho_{D-3},a}$$
or 
$$ Y_a-k_a=d h_a+\Omega _a{}^b \wedge h_b
\eqno(4.12)$$
and 
$$ \epsilon ^{\tau\lambda\rho_1\ldots \rho_{D-2}}\{ \hat
D_\lambda k_{\rho_1\ldots
\rho_{D-2}, a}-\Omega_{\lambda a}{}^c Y_{{\rho_1}\ldots \rho_{D-2},c}
-M_{\lambda\rho_1\ldots
\rho_{D-2},a}\}=0,
$$
$$\ {\rm or } \ d k_a +\Omega _a{}^b k_b 
-\Omega _a{}^b Y_b -M_a=0
\eqno(4.13)$$
where 
$$\hat D_\lambda h_{\rho_1\ldots \rho_{D-3},a}=\partial_\lambda 
h_{\rho_1\ldots \rho_{D-3},a}+\Omega_{\lambda a}{}^b h_{\rho_1\ldots
\rho_{D-3},b},
\eqno(4.14)$$
  and 
$$\Omega_{\lambda a}{}^b=(e^{-1}\partial_\lambda e)_a{}^b, \ 
h_a=h_{\rho_1\ldots \rho_{D-3},a}dx^{\rho_1}\wedge \ldots
\wedge dx^{\rho_{D-3}},\ {\rm and}\  k_a=k_{\rho_1\ldots
\rho_{D-2},a}dx^{\rho_1}\wedge \ldots
\wedge dx^{\rho_{D-2}} .
\eqno(4.15)$$ 
\par
Differentiating equation (4.12) with respect to $\partial_\nu$ and using
equation (4.13) we find the usual equations of motion of gravity.  
The calculation is most easily carried out using the forms and 
the relation 
$$d \Omega _a{}^b + \Omega _a{}^c \wedge \Omega _c{}^b =0
\eqno(4.16)$$  
Hence,   equations (4.12) and (4.13) imply Einstein's
equations of general relativity and constitute an 
interacting dual theory of gravity in terms of equations which are first
order in time derivatives. In fact,  one can also use equations 
(4.12) and
(4.13) with the $\Omega _a{}^b$ terms set to zero and  find
the same result. 
\par
Let us now consider a non-linear realisation of very extended $A_{D-3}$,
but  including only fields of level zero and one. We  therefore have the
generators
$$K^a{}_b,  R^{a_1\ldots a_8,b},
\eqno(4.17)$$ 
We will take the  local subgroup to be 
the Chevalley invariant subgroup   and so the group element takes the
form 
$$g=e^{x^\mu P_\mu} e^{h_a{}^b K^a{}_b}
e^{\hat h_{a_1\ldots a_{D-3},b}  R^{a_1\ldots a_{D-3},b}}
\eqno(4.18)$$ 
where $\hat h_{[a_1\ldots a_{D-3},b]}=0$. The Cartan forms are given by 
$${\cal V}=g^{-1}d g-\omega= dx^\mu ( e_\mu{}^aP_a+\Omega_\mu{}_a{}^b
K^a{}_b +\tilde D_\mu \hat h_{a_1\ldots a_{D-3},b}R^{a_1\ldots a_{D-3},b})
-\omega
\eqno(4.19)$$
where $\omega={1\over 2} dx^\mu \omega _\mu{}_{bc} J^{bc}$ and 
$$ \tilde D_\mu \hat h_{a_1\ldots a_{D-3},b}=\partial_\mu \hat
h_{a_1\ldots a_{D-3},b}+ \Omega_\mu{}_{a_{1}}{}^c \hat h_{c\ldots
a_{D-3},b}+\ldots  +\Omega_\mu{}_{b}{}^c \hat h_{a_{1}\ldots a_{D-3},c}
\eqno(4.20)$$
We treat  the Lorentz part of the subgroup in a preferred manner by
including in
$g$ all of $K^a{}_b$ and as a result  introducing  the above   spin
connection. As discussed in reference [9], one can solve for the spin
connection in terms of the Cartan form $\Omega_\mu{}_a{}^b$ which  a
way that is    uniquely determined  by conformal invariance.  
\par
Examining equations (4.12) and (4.13) when written in terms of tangent
indices we find that they are indeed formulated in terms of the above
Cartan forms, namely $e_\mu{}^a$, $\Omega_c{}_a{}^b$, or 
$\omega_c{}_a{}^b$, and   
$$\tilde D_{a_1} \hat h_{a_2\ldots a_{D-2},b}=e_{a_1}{}^{\mu_1}\ldots 
e_{a_{D-2}}{}^{\mu_{D-2}}(\partial_{\mu_1}\hat  h_{\mu_2\ldots
\mu_{D-2},b}+
\Omega _b{}^c \hat h_{\mu_2\ldots \mu_{D-2},b})
,$$
 plus the fields
$k_{a_1\ldots a_{D-2},b}$ and
$h_{[a_1\ldots a_{D-3},b]}$. 
Clearly, describing gravity by a non-linear realisation based on the
group very extended $A_{D-3}$ with its subgroup taken to be the Cartan
involution invariant subgroup implies, modulo a miracle involving the
inverse Higgs mechanism,  that one would have an infinite number of fields
or Goldstone bosons. Since we do not possess such a system at present we
can at best hope to find a system that has the fields of levels zero and
one and  is invariant under the action of the generators of very extended 
$A_{D-3}$ up to level one. In the above, we have found a system of
equations that is equivalent to Einstein's general relativity, 
involves the required fields and is invariant under the Borel subgroup of 
 very extended 
$A_{D-3}$ up to level one. It is clearly invariant under the Lorentz
group, but it remains to show that it is invariant under the remaining
generator at level one. 
The fields $k_{a_1\ldots a_{D-2},b}$ and
$h_{[a_1\ldots a_{D-3},b]}$ play the role of the infinite number of
fields occurring at levels two and above that we must add in order to have 
the full non-linear realisation. 
\par
Of course, given the Cartan forms of equation (4.19)
 one would not immediately conclude that the required equations of
motion were those of equations (4.12) and (4.13). However, as explained in
reference [9] one must demand simultaneous invariance under the conformal
group. Carrying out this requirement should result in equations
(4.12) and (4.13) being the unique equations invariant under both
algebras up to the level being considered.  This, and its extension
to the next level of very extended $A_8$,  would make  the calculation
given in  this section more compelling.  
\par
We close this section by generalising the above formulation of gravity to 
include matter and in particular consider a rank $p$ gauge field
$A_{(p)}=A_{\mu_1\ldots \mu_p} dx^{\mu_1}\wedge \ldots \wedge
dx^{\mu_p}$ with  field strength $F_{(p+1)}=(p+1)d A_{(p)}$. The dual
field strength being given by $G_{(D-p-1)}=\star F_{(p+1)}$. To
incorporate matter we define 
$$\hat M_a=M_a+{1\over (D-p-1)}G_{(D-p-1)}\wedge F_{(p)a}
-(-1)^p {1\over (p+1)}G_{(D-p-2)a}\wedge F_{(p+1)})
\eqno(4.21)$$ 
where $ F_{(p)a}=F_{\mu_1\ldots \mu_p a} dx^{\mu_1}\wedge \ldots \wedge
dx^{\mu_p}$ and similarly for $G_{(D-p-2)a}$. If we replace $M_a$ by
$\hat M_a$ in equation (4.6) or equation   (4.13) we find that 
$$
R_\mu{}^d-{1\over 2}e_\mu{}^d R= c\{ F_{\mu\lambda_1\ldots \lambda_p}
F^{d\lambda_1\ldots \lambda_p}-{1\over 2(p+1)}e_\mu{}^d 
 F_{\lambda_1\ldots \lambda_{p+1}}
F^{\lambda_1\ldots \lambda_{p+1}} \}
\eqno(4.22)$$
where $c$ is a constant which can be adjusted by rescaling $A_{(p)}$.
We recognise the right hand side as the required contribution to the
energy-momentum tensor. As must be the case for consistency of the
equations of motion, and 
as is required by the 
conservation of energy and momentum, one can verify that $d\hat M_a=0$. 
\par
We observed that the above formulation of gravity involves  the
field $ h_{a_1\ldots a_{D-3},b}$ which is not subject to a constraint 
$h_{[a_1\ldots a_{D-3},b]}=0$.  
 If we suppose that the equation to describe
the dual formulation of gravity must be first order in space-time
derivatives; then,  at the linearised level, it must  be of the form 
$$ \epsilon ^{\mu\nu\tau_1\ldots \tau_{D-2}}
\partial_{\tau_1} h_{\tau_2\ldots \tau_{D-2},}{}^d+\ldots =f\partial^\mu
h^\nu{}^d +\ldots
\eqno(4.23)$$
where $f$ is a constant and $+\ldots$ stands for all possible terms linear
in the space-time derivatives and in the same field, but with different
index structure. Furthermore,  one must be able to apply a derivative to
the left hand-side of the above equation and be able to eliminate the
field
$h_{a_1\ldots a_{D-3},b}$ and obtain an equation in terms of $h_a{}^b$
alone  which is the linearised Einstein equation. 
 Writing down the most general
terms and imposing the constraint $h_{[a_1\ldots a_{D-3},b]}=0$ we find
constraints on the coefficients that are not compatible with deriving 
the linearised Einstein equation from equation (4.3). However, equations
(4.1) and (4.2), at the linearised level, do provide  a system of
equations which  lead to the linearised Einstein equation,
but they involve the unconstrained field
$h_{a_1\ldots a_{D-3},b}$. 
\par
It is instructive to discuss the dual formulation of gravity given above
with those given in references [15-17]. In these references 
 a dual formulation of gravity is given at the linearised
level    by an equation which is second
order in space-time derivatives  of the form
$$R_{\mu\nu}{}^{ab}=
\epsilon _{\mu\nu}{}^{\rho_1\ldots\rho_{D-2}}(\partial_a\partial_{[\rho_1}
l_{\rho_2\ldots \rho_{D-2}], b}-(a\to b)),
\eqno(4.24)$$ 
where $l_{[a_1\ldots a_{D-3},b]}=0$. 
Using equation (4.1) one can calculate $R_{\mu\nu}{}^{ab}$ in terms of 
$Y_{\tau_1\ldots\tau_{D-2},} {}_d$ and, after a field redefinition, it can
be brought to the form of equation (4.24).  This can be seen to be a
consequence of the fact that an  additional space-time
derivative allows an additional  gauge symmetry which can be used to
algebraically gauge it away, or equivalently, eliminate its appearance in
the appropriate variables.  This gauge symmetry is 
related to the "local Lorentz" transformation discussed below
equation (4.3). Hence, from this view point the field
$h_{[a_1\ldots a_{D-3},b]}$ is not really present. However, to place
gravity on the same footing as the gauge fields we would like to have a
dual formulation of gravity that is expressed by equations that are first
order in  space-time derivatives.  The gauge field $A_{a_1\ldots a_3}$
requires only one additional field $A_{a_1\ldots a_6}$ to rewrite its
equations in terms of equations involving only one space-time derivative.
However,  the gravitational equations are essentially non-linear
equations and this might be viewed as the source of the infinite number of
fields required by the conjectured very extended $A_8$ symmetry. It would
be interesting to see how the fields at the next level of 
very extended $A_8$ symmetry replace $h_{[a_1\ldots a_{D-3},b]}$  and the
other additional field $k_{[a_1\ldots a_{D-2},b]}$ even at the
linearised level.  
\medskip
{\bf {5 Eleven Dimensional Supergravity and $E_{11}$}}
\medskip
In this section we sketch how, starting from $E_{11}$, we can find eleven
dimensional supergravity as a non-linear realisation. This contrasts with 
reference [2] where we started with the non-linear realisation of
reference [9], identified $E_{11}$, and then found substantial fragments
of an $E_{11}$ symmetry.  Most of the equations given below can be found
in  reference [2], but it is encouraging to see them
emerge starting from
$E_{11}$. In section three we found that the first
three levels of
$E_{11}$  contain the generators in  equations (3.5-3.9) plus the
generators associated  with negative roots given in equation (3.10). 
 In a non-linear realisation all generators in
the symmetry algebra not in the local subgroup must correspond to fields
in the theory. The only exception to this is when one can use the so
called  inverse Higgs effect, but  this mechanism will not concern us
here. As explained in references [9,2], the level one and two generators 
$R^{a_1a_2a_3}$  and
$R^{a_1\ldots a_6}$ correspond to the gauge fields  $A_{a_1a_2a_3}$ 
 and its dual
$A_{a_1\ldots a_6}$ respectively. Gravity can be described by the 
the group $GL(11)$ and so corresponds to the level zero generators
$K^a{}_b$ associated with the fields $h^a{}_b$. However, as realised
in reference [2], demanding that 
$E_8$ and $A_{10}$ be symmetry groups  implies that we should use a
dual formulation of gravity which  involves the field
$\hat h_{a_1\ldots a_8,b}$ corresponding to the generator $R^{a_1\ldots
a_8,b}$. Thus, there is a very close correspondence between the fields
introduced to describe eleven dimensional supergravity as a non-linear
realisation and  the fields content up to level three required by
demanding $E_{11}$ as a symmetry group. This can be taken as yet another
encouraging sign that the maximal supergravity theories really are
invariant under $E_{11}$.  
\par
Taking the Cartan involution invariant subgroup as the local subgroup, up
to level three, the non-linear realisation is  constructed from
the group element [9,2] 
$$g=e^{h_a{}^b K^a{}_b} e^{{1\over 3!} A_{a_1\ldots a_3}R^{a_1\ldots a_3}}
e^{{1\over 6!} A_{a_1\ldots a_6}R^{a_1\ldots a_6}} e^{\hat h_{a_1\ldots
a_{D-3},b} R^{a_1\ldots a_{D-3},b}}
\eqno(5.1)$$ 
The equations of motion should be constructed from the Cartan forms 
${\cal V}=g^{-1}d g-w$ and they are given by [9,2]
$${\cal V}= dx^\mu(e_\mu{}^a  P_a + \Omega _{\mu a}{}^b K^a{}_b
+ {1\over 3!}\tilde D_\mu A_{c_1\ldots c_3} R^{c_1\ldots c_3}
+{1\over 6!}\tilde D_\mu A_{c_1\ldots c_6} R^{c_1\ldots c_6}+
\tilde D_\mu\hat h_{c_1\ldots c_8,b} R^{c_1\ldots c_8,b})-\omega
\eqno(5.2)$$
where 
$$e_\mu{}^a \equiv (e^h)_\mu{}^a, \ \tilde D_\mu A_{c_1\ldots c_3}\equiv  
 \partial_{\mu} A_{c_1 c_2 c_3} 
+ ((e^{-1}\partial _\mu e)_{c_1}{}^{b}A_{b c_2 c_3}+ \dots ), 
$$
$$ 
\tilde  D_\mu A_{c_1\ldots c_6}\equiv  
\partial _\mu A_{c_1\ldots c_6}+ 
((e^{-1}\partial _\mu e)_{c_1}{}^{b}A_{b c_2 \ldots c_6}+ \dots )
- 20(A_{[ c_1\ldots c_3}\tilde D_\mu A_{c_4\ldots c_6]})
$$
$$
\tilde D_\mu\hat  h_{c_1\ldots c_8,b} =
\partial_\mu \hat h_{c_1\ldots c_8,b}+( (e^{-1}\partial _\mu e)_{c_1}{}^d
\hat h_{d c_2\ldots c_8,b} +\dots)
$$
$$-{1\over (3!)^3}A_{[c_1\ldots c_3}\tilde D_\mu A_{c_4c_5c_6}
A_{c_7c_8 ]b}-
{1\over (3!)(6!)}A_{[c_1\ldots c_6}\tilde D_\mu A_{c_7c_8 ]b}-tr
\eqno(5.3)$$
where $+\ldots $ denotes the action of 
$(e^{-1}\partial _\mu e)$ on the other 
indices and $-tr$ means that one should subtract a term in order to
ensure that $\tilde D_\mu h_{[c_1\ldots c_8,b]}=0$.  
\par
As explained in references [2,9] we require invariance not only under
$E_{11}$ but also under the conformal group. This implies that  we should
use  only those  combinations of $E_{11}$ Cartan forms which are covariant
under the latter group.  The result of this procedure for the rank three 
and six forms is 
that one should only use the simultaneously covariant forms [2,9]
$$\tilde F_{c_1\ldots c_4}
\equiv  4(e_{[ c_1}{}^\mu \partial _\mu A_{c_2\ldots c_4]}
+ e_{[ c_1}{}^\mu 
( e^{-1} \partial_\mu  e)_{ c_2 }{\ \  }^b  
A_{b c_3 c_4]}+\ldots)
\eqno(5.4)$$
and 
$$
\tilde F_{c_1\ldots c_7}\equiv  
7(e_{[ c_1}{}^\mu(\partial _\mu A_{c_2\ldots c_7]})
+ e_{[ c_1}{}^\mu( e^{-1} \partial_\mu  e)_{ c_2 }{\ \ \ }^b A_{b
c_3 \ldots c_7]}  +\ldots  +5 \tilde F_{[c_1\ldots c_4}\tilde
F_{c_5\ldots c_7]}) 
\eqno(5.5)$$
The invariant  equation of motion is then given by 
$$
\tilde F^{c_1\ldots c_4}={1\over 7!}
\epsilon _{c_1\ldots c_{11}}\tilde F^{c_5\ldots c_{11}}
\eqno(5.6)$$
\par
There only remains the equation for gravity. This equation should be given
by equation (4.12) but with $\hat D_\mu h_{c_1\ldots c_8,b} $ of equation
(4.14) replaced by $\tilde D_\mu \hat h_{c_1\ldots c_8,b}$ of equation
(5.3) and the addition of the field $h_{[c_1\ldots c_8,b]}$.  Taking $d$
of this equation one does indeed find terms similar to those on the
right-hand side  of equation (4.21) which are   those  required to find
the energy momentum tensor of the four form field strength on the
right-hand side of the Einstein equation.  However, one also finds
expressions which are not written in terms of the four form field
strength or its dual and these should be canceled by terms added to the
definitions of $Y_a$ and $M_a$. We hope to return to this calculation in
the future. 
\par
The level in $E_{11}$ is measured by the generator ${1\over 3}D= {1\over
3}\sum_a K^a{}_a$ where $K^a{}_b$ are the generators of GL(11). Clearly, 
$D$ commutes with the Cartan subalgebra generators  and simple roots of
SL(11), but with the simple root generator $E_c=R^{91011}$ one has the
commutator
$[D, E_c]=E_c$ as required. However, the generator $D$ corresponds to a
symmetry of eleven dimensional supergravity acting with $e^{rD}$ on the
group element of equation (5.1) one finds that $x^\mu$ scales and
$h_a{}^b\to h_a{}^b+rh_a{}^b$ with the other fields being inert. It is
then easy to see that  the Cartan forms referred to the tangent
space, such as in equation (5.4) and (5.5), are  inert under
$D$ transformations. 
As such,   the $D$ symmetry  is automatically
preserved by the equations of motion  constructed from the Cartan
forms. In fact,  in a Lorentz  covariant formulation the level of the
fields does not appear to  provided a very  useful means ordering the
calculation.    This is
apparent from equation (2.14) which relates fields of different
levels, proceeding level by level implies that at the first level the
right-hand side of equation (5.7) vanishes. 
\medskip
{\bf {Acknowledgment}}
The author wishes to thank David Olive, Marc Henneaux, Lionel Mason,
Boris Pioline and Igor Schnakenburg for discussions. 
\medskip
\medskip
{\bf {References}}
\medskip
\item{[1]} E. Cremmer, B. Julia and J. Scherk, Phys. Lett. 76B
(1978) 409.
\item{[2]} P. West, {\it $E_{11}$ and M Theory}, Class. Quant. Grav. 18
(2001) 4443 , hep-th/0104081. 
\item {[3]} B. de Wit and H. Nicolai, Nucl. Phys. B274 (1986)
363; H.  Nicolai Phys. Lett. 155B (1985) 47; 
\item{[4]}  H.  Nicolai, Phys. Lett. 187B (1987) 316    
\item{[5]} S. Melosch and H. Nicolai, {\it ``New Canonical
Variables for $D=11$ Supergravity"}, hep-th/9709227; 
 K. Koespell, H. Nicolai and H. Samtleben, {\it ``An Exceptional 
Geometry for $d=11$ Supergravity"}, Class. Quantum Grav. 17 (2000)
3689.
\item{[6]} G. Moore, {\it `` Finite in all directions"},
hep-th/9305139; 
 P. West, {\it ``Physical States and String Symmetries"},
hep-th/9411029,  Int.J.Mod.Phys. {\bf A10} (1995) 761. hep-th/9411029.
\item{[7]} J. Harvey and G. Moore, {\it On the Algebra of BPS States},
Nucl. Phys. B463 (1996) 315, hep-th/9510182, Exact Gravitational
Threshold Corrections in the FHSV model, hep-th/9611176. 
\item{[8]} T. Damour, M. Henneaux, {\it ``E(10), BE(10) and
arithmetical chaos in superstring cosmology ''}, Phys. Rev.Lett. {\bf
86} (2001) 4749, {\tt hep-th/0012172};  
 T. Damour, M. Henneaux, B. Julia and H. Nicolai, {\it Hyperbolic
Kac-Moody Algebras and Chaos in Kaluza-Klein Models},
hep-th/0103094.  
\item{[9]}  P.~C. West, {\it ``Hidden superconformal symmetry in {M}
    theory ''},  JHEP {\bf 08} (2000) 007, {\tt hep-th/0005270}
\item{[10]} I. Schnakenburg and  P. West, {\it ``Kac-Moody 
Symmetries of IIB Supergravity,''}, Phys. Lett. {\bf B517} (2001)
421, {\tt hep-th/0107181}
\item{[11]} N.Lambert and  P. West, {\it ``Coset symmetries in
dimensionally reduced bosonic string theory''}, Nucl.Phys. {\bf B615}
117, {\tt hep-th/0107209}
\item{[12]} T. Damour, M.  Henneaux and H. Nicolai, {\it $E_{10}$ and a
small tension expansion of M theory}, hep-th?0207267.
\item{[13]} D. Olive, M. Gaberdiel and P. West,  {\it A Class
of Lorentzian Kac-Moody Algebras} hep-th/0205068, Nucl.Phys. to be
published  
\item{[14]} V. Kac, {\it ``Infinite Dimensional Lie
Algebras"}, chapter 5, Birkhauser, 1983. 
\item{[15]} T. Curtright, {\it Generalised Gauge Fields}, Phys. Lett. B165
(1985) 304; C. Aulakh, I. Koh and S. Ouvry, Phys. Lett. B173 (1986) 284. 
\item{[16]} C. Hull, {\it Duality in Gravity and Higher Spin Gauge
fields}, JHEP09 (2001) 027, hep-th/0107149. 
\item{[17]} X. Bekaert and N. Boulanger, {\it Tensor gauge Fields in
Arbitrary Representations of GL(D, R): duality and Poincare Lemma},
hep-th/0208058; X. Bekaert, N. Boulanger and M. Henneaux,{\it Consisitent
Deformations of Dual Formulations of Linearised Gravity: a no go theorem} 
hep-th/0210278.
\item{[18]} G. Sparling, {\it Twistor, Spinors and the Einstein Vacuum
Equations}, Advances in Twistor Theory, Vol III, Curved Twistor Spaces, 
p 179-186, ed. L. J. Mason, L. P. Hughston, P. Z. Kobak, K. Pulverer, 
Research notes in Mathematics 424, (2001), Chapman and Hall/CRC, 
originally appeared in  a Twistor News Letter in  (1982).
\item{[19]} I thank Boris Pioline for explaining to me how to do this
using the paper, N. Obers and B. Pioline, {\it U-duality and M theory, an
algebraic approach}, hep-th/9812139.
\item{[20]} I wish to thank Hermann Nicolai for pointing out that the 
original version of this paper omitted the $p_1=1$ case from equation (3.3).

\end